\documentstyle[aps,preprint]{revtex}
\tightenlines 
\begin{document}
\draft

\title{Theory of bound polarons in oxide compounds}

\author{J. T. Devreese\cite{A1}, V. M. Fomin\cite{A2}}
\address{Theoretische Fysica van de Vaste Stof,
Universiteit Antwerpen (U.I.A.),
Universiteitsplein 1, B-2610 Antwerpen, Belgium}
\author{E. P. Pokatilov}
\address{Laboratory of Multilayer Structure Physics,
Department of Theoretical Physics,
State University of Moldova,
Strada A. Mateevici, 60,
MD-2009 Kishinev, Moldova}
\author{E. A. Kotomin\cite{A3}, R.~Eglitis}
\address{Fachbereich Physik, Universit\"at Osnabr\"uck, D-49069
Osnabr\"uck, Germany}
\author{Yu. F. Zhukovskii}
\address{Institute of Solid State Physics, University of Latvia,
Kengaraga 8, LV-1063 Riga, Latvia}
\maketitle

\bigskip
\bigskip
\small
We present a multilateral theoretical study of bound polarons in
oxide compounds MgO and  $\alpha$-Al$_2$O$_3$ (corundum). A continuum theory
at arbitrary electron-phonon coupling is used for calculation
of the energies of thermal dissociation, photoionization (optically induced
release of an electron (hole) from the ground
self-consistent state), as well as optical absorption to
the non-relaxed excited states. Unlike the
case of free strong-coupling polarons, where the ratio $\kappa$ of the
photoionization energy to the thermal dissociation energy was shown to be
always equal to 3, here this ratio depends on the
Fr\"ohlich coupling constant $\alpha$ and the screened Coulomb
interaction strength $\beta$. Reasonable variation of these two parameters
has demonstrated that the magnitude of $\kappa$ remains usually in the narrow
interval from 1 to 2.5.
This is in agreement with atomistic calculations
and experimental data for hole O$^-$  polarons bound to the cation vacancy in
MgO.
The thermal dissociation energy for the ground self-consistent
state and the energy of the optically induced charge transfer process
(hops of a hole between O$^{2-}$ ions) have been calculated using the
quantum-chemical method INDO. Results obtained
within the two approaches for hole O$^-$ polarons bound by the
cation vacancies (V$^-$) in MgO  and by the Mg$^{2+}$ impurity (V$_{Mg}$)
in corundum are compared to experimental data and to each other. We discuss a
surprising closeness of the results obtained on the basis of independent models
and their agreement with experiment.
\normalsize
\pacs{PACS numbers: {73.}{61.Ng},{71.}{38.+i},{71.}{15.$-$m}}

Properties of polarons and bipolarons in oxide materials
continue to attract considerable attention in solid state and materials
physics due to their possible relevance to  the high-T$_c$
superconductivity \cite{1}. Theory for large-radius \cite{2} and
small-radius \cite{3} polarons is well developed.
At moderate temperatures polarons are usually bound to
defects (vacancies or impurities)
unavoidably present in any material. Examples  are a hole
trapped by the cation vacancy in MgO (called V$^-$ center)
and a hole
trapped by the Mg$^{2+}$ impurity substituting for a regular Al$^{3+}$
atom in Al$_2$O$_3$ (called V$_{Mg}$ center) \cite{4}.

Fundamental characteristics of free polarons are energies of their
photoionization ($E_{\rm opt}^f$) and thermal dissociation ($E_{\rm th}^f$).
From the well-known
theorem 1:2:3:4 derived by Pekar \cite{5} for relations between the
average kinetic energy of the electron,
the polarization energy of the crystal $E_{\rm polariz}^f$,
the modulus of the electron eigenenergy $E_{e}^f$,
and the average potential energy of
the interaction between the electron and the polarized crystal in the
strong-coupling polarons, it follows, in particular, that the ratio
between the photoionization energy $E_{\rm opt}^f=\vert E_{e}^f\vert$
and the thermal dissociation energy $E_{\rm th}^f=\vert E_e^f\vert - E_{\rm polariz}^f$
for a free polaron $\kappa^f \equiv E_{\rm opt}^f/E_{\rm th}^f
=\vert E_{e}^f\vert/(\vert E_e^f\vert - E_{\rm polariz}^f)$
equals exactly 3. A generalization of this ground-state theorem, which
relates the interaction energy to the kinetic energy for free
polarons, was derived by Lemmens, De Sitter and Devreese for all values of the
electron-phonon coupling constant \cite{L73}. Optical
transitions of large polarons with taking relaxed excited states into account
were examined in Refs. \cite{E65}.

Calculation of the ratio $\kappa = E_{\rm opt}/E_{\rm th}$ of the
photoionization and thermal dissociation energies
allows one to examine the adequacy of different theoretical approaches
to bound polarons.
In the present Letter we analyze this problem in two ways:
using an extended {\it continuum} theory of large radius {\it bound} polarons,
and performing {\it atomistic} calculations for the particular
hole polarons in the above-mentioned ionic solids --- MgO and corundum
($\alpha$-Al$_2$O$_3$). These two types of calculations allow us to draw
conclusions how the continuum approach can describe models, where the
lattice polarization is of Fr\"ohlich type, while the state of charge carriers
is of small radius (such a model has been suggested e.\,g. in Ref. \cite{A99}
for a small Fr\"ohlich polaron). In the framework of this approach,
unknown values of the macroscopic parameters should be obtained from
comparison with experimental data. At the same time, we examine the
efficiency of the microscopic quantum-mechanical approaches for solving
the aforementioned problems of bound states.

A variational theory of bound polaron states which embraces the overall
interval of values of the Fr\"ohlich coupling constant
$\alpha$ and of the strength $\beta$ of the Coulomb interaction between
a band charge carrier and a charged impurity was developed in Ref. \cite{6}.
The Hamiltonian of an electron (hole) in the field of an
impurity interacting with the polar longitudinal optical (LO) phonons is
\begin{eqnarray}
H=\frac{{\bf p}^{2}}{2m_b} -{e^2\over \varepsilon_0 r}  + \sum_{{\bf k}}
\hbar \omega_{\mbox{\tiny LO}} a^{\dagger}_{{\bf k}}  a_{{\bf k}}
+\sum_{{\bf k}} (V_{k}   a_{{\bf k}}  {\rm e}^{i {\bf {\bf k} \cdot r}} + {\mbox{h.c.}}),
\label{1a}\end{eqnarray}
where ${\bf  r}$ is the position coordinate operator of the electron (hole)
with band mass $m_b$, ${\bf p}$ is its canonically conjugate momentum
operator; $a^{\dagger}_{{\bf k}}$ and $a_{{\bf k}}$ are the creation (and annihilation)
operators for LO phonons of wave vector ${\bf k}$ and
frequency $\omega_{\mbox{\tiny LO}}$. The $V_{k}$ are Fourier components of the
electron(hole)-phonon interaction
\begin{eqnarray}
V_{k}=-i \frac{\hbar \omega_{\mbox{\tiny LO}}}{k}
\left( \frac{4\pi \alpha}{V}        \right)^{\frac{1}{2}}
\left( \frac{\hbar}{2m_b\omega_{\mbox{\tiny LO}}}   \right)^{\frac{1}{4}}
\label{1b}\end{eqnarray}
with the Fr\"ohlich coupling constant
\begin{eqnarray}
\alpha = {1 \over \hbar \omega_{\mbox{\tiny LO}}} {e^2 \over
\sqrt{2}}
\left({1\over \varepsilon_{\infty}} - {1\over \varepsilon_0}\right)
\sqrt{{m_b\omega_{\mbox{\tiny LO}}\over \hbar}},
\label{1bb}\end{eqnarray}
where $ \varepsilon_0 $ and $ \varepsilon_{\infty} $ are the static and
high-frequency dielectric constants of the crystal, respectively.
Choosing the Feynman units, when the polaron radius
$\sqrt{\hbar / (2m_b\omega_{\mbox{\tiny LO}})}$ and the phonon energy
$\hbar \omega_{\mbox{\tiny LO}}$ serve as units of length and energy,
respectively,
we represent the second term in the right-hand side of Eq. (\ref{1a})
in the form
$-\beta / r$, where the dimensionless Coulomb interaction strength
relates to the Fr\"ohlich coupling constant (\ref{1bb}) as
follows:
\begin{eqnarray}
\beta = 2\alpha/( {\varepsilon_0} / \varepsilon_{\infty} - 1).
\label{1}
\end{eqnarray}

The approach, developed by Devreese {\it et al.} \cite{6},
consists in two stages. First, the unitary transformation is performed
\begin{eqnarray}
U = \exp\left[ \sum_{\bf k} \left(V_{k} a_{\bf k} \rho_{\bf k} - {\mbox{h.c.}}\right)\right],
\quad \rho_{\bf k} = \langle \phi_n\vert
{\rm e}^{i {\bf {\bf k} \cdot r}} \vert \phi_n\rangle,
\label{2a}\end{eqnarray}
where $\vert \phi_n \rangle$ is the electron (hole) wave function,
which will be found variationally. This transformation of the Hamiltonian
(\ref{1a}) results in
\begin{eqnarray}
&&H' \equiv U^{-1}HU =  H_0+
\sum_{{\bf k}} a^{\dagger}_{\bf k}  a_{\bf k} + H_{int},\\
&&H_0={\bf p}^{2} -{\beta \over r}  + \sum_{{\bf k}}
\vert V_{k}\vert ^2  \vert \rho_{\bf k}\vert ^2
-\sum_{{\bf k}} \vert V_{k}\vert ^2
(\rho^*_{\bf k}{\rm e}^{i {\bf {\bf k} \cdot r}} + {\mbox{c.c.}}),\\
&&H_{int}= \sum_{\bf k} (V_{k}   a_{\bf k}  ({\rm e}^{i {\bf {\bf k} \cdot r}}-
\rho_{\bf k}) + {\mbox{h.c.}}).
\label{2}\end{eqnarray}
Second, the following trial function is proposed in Ref.\,\cite{6},
which is proved to be a fair approximation
for all values of $\alpha$ and $\beta$:
\begin{eqnarray}
\vert \psi\rangle =c\vert 0\rangle \vert\phi_n \rangle +
\sum_{{\bf k}}  V^*_{k} g^*_{\bf k}({\rm e}^{-i {\bf {\bf k} \cdot r}}-
\rho^*_{k}) \vert 0\rangle \vert \phi_n\rangle,
\label{3}\end{eqnarray}
where $c$ is a normalization constant and $\vert 0\rangle$ is the phonon
ground state.
The variational principle with the trial function (\ref{3}) results in
\begin{eqnarray}
{g_{\bf k} \over c} = -{1-\vert \rho_{\bf k}\vert ^2 \over
D_1({\bf k}) + D_2({\bf k}) + (1-\vert \rho_{\bf k}\vert ^2)\chi/2},
\label{4a}\end{eqnarray}
where
\begin{eqnarray}
&&\chi = {1\over c}\sum_{\bf k} \vert V_{k}\vert ^2
(1-\vert \rho_{\bf k}\vert ^2)(g_{\bf k}+g^*_{\bf k}),\\
&&D_1({\bf k})= \langle \phi_n\vert ({\rm e}^{i {\bf {\bf k} \cdot r}}-
\rho_{\bf k}) H_0 ({\rm e}^{-i {\bf {\bf k} \cdot r}}-
\rho^*_{\bf k}) \vert \phi_n\rangle,\\
&&D_2({\bf k})=(1-\vert \rho_{\bf k}\vert ^2) (1-\langle \phi_n\vert
H_0 \vert \phi_n\rangle).
\label{4b}\end{eqnarray}
The variational energy for the bound polaron is found to be
\begin{eqnarray}
E =  \langle \phi_n\vert H_0 \vert \phi_n\rangle + {\chi \over 2}.
\label{5}\end{eqnarray}
This energy is to be minimized with respect to the parameters entering the
trial function $\vert \phi_n\rangle$. It is this minimal energy which
relates to the thermal dissociation of a polaron bound to an impurity:
$E _{\rm th} \equiv \vert  E_{\rm min} \vert$. The electron (hole) eigenenergy is
$E_c =  E_{\rm min} - E_{\rm polariz}$, where $E_{\rm polariz}$ is the
elastic energy of the polarization (i.e., the
energy of the ionic polarization due to a lattice energy gain after
self-consistent relaxation). The minimal energy, which is absorbed during the
photoionization of a bound polaron (the optically induced release of an
electron (hole) from the ground state), equals the
modulus of the electron (hole) eigenenergy $\vert E_c \vert$:
\begin{eqnarray}
E_{\rm opt}\equiv \vert E_c \vert=E_{\rm th}+E_{\rm polariz}.
\label{15new}\end{eqnarray}
Hence, the ratio of the photoionization energy and the thermal dissociation
energy is
\begin{eqnarray}
\kappa \equiv { E_{\rm opt} \over E_{\rm th}} = { E_{\rm th}+E_{\rm polariz} \over E_{\rm th}}.
\label{4c}\end{eqnarray}

In order to find the ratio $\kappa $, it is necessary to calculate
the eigenenergy and the energy of the polarization field.
Under the unitary transformation of Eq. ({\ref{2a}),
the Hamiltonian of the phonon subsystem turns to
\begin{eqnarray}
H'_{ph} \equiv U^{-1}  \sum_{{\bf k}} a^{\dagger}_{\bf k}  a_{\bf k}  U =
\sum_{\bf k} a^{\dagger}_{\bf k}  a_{\bf k}+
\sum_{\bf k} \vert V_{k}\vert ^2 \vert \rho_{\bf k}\vert ^2 -
\sum_{\bf k} (V_{k} \rho_{\bf k} a_{\bf k}+ {\mbox{h.c.}}).
\label{6}\end{eqnarray}
Averaging the transformed Hamiltonian of the phonon subsystem on the
trial function (\ref{3}) gives the energy of the polarization field in a
polaron bound to an impurity:
\begin{eqnarray}
E_{\rm polariz}  \equiv \langle \psi \vert H'_{ph}\vert \psi \rangle =
{\lambda \over 1+ \lambda} +
\sum_{\bf k} \vert V_{k}\vert ^2 \vert \rho_{\bf k}\vert ^2
\label{7}\end{eqnarray}
with
\begin{eqnarray}
\lambda =\sum_{\bf k} {\vert V_{k}\vert ^2
(1-\vert \rho_{\bf k}\vert ^2)^3\over \left[D_1({\bf k})
+ D_2({\bf k}) - (1-\vert \rho_{\bf k}\vert ^2)\chi/2\right]^2}.
\label{7a}\end{eqnarray}
The variational parameters  which deliver a minimum to the
polaron energy, should be substituted to Eq. (\ref{7}).
In particular, for a spherically symmetrical variational function
$\vert \phi_n \rangle$ the expressions in the right-hand side of
Eq. (\ref{7}) can be simplified to
\begin{eqnarray}
&&\lambda = {2\alpha \over \pi}\int_0^{\infty} dk
{(1-\vert \rho_{k}\vert ^2)^3\over \left[D_1({k})
+ D_2({k}) - (1-\vert \rho_{k}\vert ^2)\chi/2\right]^2},\\
&& \sum_{\bf k} \vert V_{k}\vert ^2 \vert \rho_{\bf k}\vert ^2=
{2\alpha \over \pi}\int_0^{\infty} dk
\vert\rho_{k}\vert ^2.
\label{9}\end{eqnarray}
Using for the ground state the hydrogen-like function,
which was suggested in Ref. \cite{6},
\begin{eqnarray}
\vert \phi_{1s} \rangle = {a^{3/2} \over 2^{3/2} \sqrt{\pi}}
{\rm e}^{-a r/2}
\label{10}\end{eqnarray}
with a variational parameter $a$,
we obtain the ratio $\kappa (\alpha,\beta)=E_{{\rm opt}, 1s}/E_{{\rm th}, 1s}$ as a
function of the parameters $\alpha$ and $\beta$ for a bound large polaron.

Results of the numerical calculation of the ratio
$\kappa (\alpha,\beta)$
for a polaron bound to an impurity
at $0 \leq \alpha \leq 10$ and $0\leq \beta \leq 10$  are plotted in Fig. 1.
As follows from this figure,
for all $\alpha \neq 0$ the ratio $\kappa (\alpha,\beta)>1$. This
inequality
arises due to the following fact. During the photoionization process, the
electron (hole) is released from the
polarization potential well which is frozen, being adjusted to the
electron (hole) wave function in the polaron ground state, while in the case
of a thermal dissociation of the bound polaron there occurs a
disintegration of the
polarization state and, consequently, the coupling of the electron (hole) to
the polarization well weakens continuously during this process.

As seen from Fig. 1, a decrease of the Coulomb attraction parameter $\beta$
at a fixed value of $\alpha$
leads to a rise in the  ratio $\kappa (\alpha,\beta)$, except for the region
of $\beta$ less than or of the order of unity at $\alpha<8$.
In particular, at $\alpha= 10$ and $ \beta =0$,
$\kappa (\alpha,\beta)$ is as large as 2.6,
already close to but still smaller than 3, obtained in the polaron
theory in the limit $\alpha \to \infty$.
When increasing
$\alpha$, at a fixed value of $\beta$, a role of the interaction of
an electron (hole) with polarization enhances, and thus, the value of the ratio
$\kappa (\alpha,\beta)$ rises. A weak maximum in the region of small values
of $\beta\sim 1$ at $\alpha<8$
apparently reflects a transition from a two-center state of a polaron
moving around the impurity center to a one-center state, where an
electron (hole) is in a unique potential well due to the common action of the
polarization field and of the impurity center.

Further, we consider the excited state of the bound polaron in the non-relaxed
polarization well of the $1s$ ground state of Eq.\,(\ref{5}), which will be referred to
as a non-relaxed excited state. The factors $V_{k}^* \rho_{\bf k}^*$
at $a_{\bf k}^{\dag}$ and $V_{k} \rho_{\bf k}$ at $a_{\bf k}$ in the unitary transformation
of Eq.\,(\ref{2a}) have the meaning of shifts of the polarization vibrations
due to the interaction of the polarization field with the electron in the state
$\varphi_n$:
\begin{eqnarray}
U^{-1}a_{\bf k}U= a_{\bf k} - V_{k}^* \rho_{\bf k}^*;\>\>\>
U^{-1}a_{\bf k}^{\dag}U = a_{\bf k}^{\dag} - V_{k} \rho_{\bf k}.
\label{UU}\end{eqnarray}
Hence, the wave function of the non-relaxed excited state $2p$ in the frozen polarization
well correlated to the electron state $1s$ has the form:
$$
\vert \psi_{2p}\rangle =c\vert 0\rangle \vert\phi_{2p} \rangle +
\sum_{{\bf k}}  V^*_{k} g^*_{\bf k}A_{\bf k}\vert 0\rangle \vert \phi_{2p}\rangle,\>\>\>
A_{\bf k}={\rm e}^{-i {\bf {\bf k} \cdot r}}-\rho^{*1s}_{\bf k},
$$
where $\rho_{\bf k}^{1s} = \langle \phi_{1s}\vert
{\rm e}^{i {\bf {\bf k} \cdot r}}\vert \phi_{1s}\rangle$. Like in the case of
self-consistent polaron states, $g_{\bf k}$ is a variational function. After
transformations similar to the above-described ones we obtain the following
variational energy of the non-relaxed excited $2p$ state:
\begin{eqnarray}
E_{2p}^{1s} =  \langle \phi_{2p}\vert H_0 \vert \phi_{2p}\rangle +
{\chi_{2p}^{1s} \over 2},
\label{EE}\end{eqnarray}
where in
$$
\vert \phi_{2p} ({\bf r})\rangle = {b^{5/2} \over 2^{5/2} \sqrt{\pi}}
{\rm e}^{-b r/2}r \cos{\theta}
$$
$b$ is a variational parameter and $\theta$ is the angle between ${\bf r}$
and the $z$-axis. In Eq. (\ref{EE}),
\begin{eqnarray}
&&\chi^{1s}_{2p} =-{\alpha \over \pi}\int_0^{\infty} dk \int_0^{\pi} \sin\theta
d\theta
{\langle A_{\bf k}^* A_{\bf k}\rangle^2 \over
D_1^{1s}({\bf k})+D_2^{1s}({\bf k}) -
\langle A_{\bf k}^* A_{\bf k}\rangle \chi^{1s}_{2p}/2}, \\
&&D_1({\bf k})= \langle \phi_{2p}\vert A_{\bf k}  H_0 A_{\bf k}^*
\vert \phi_{2p}\rangle,\\
&&D_2({\bf k})= \langle A_{\bf k}^* A_{\bf k}\rangle
(1-\langle \phi_{2p}\vert
H_0 \vert \phi_{2p}\rangle).
\label{CHI}\end{eqnarray}
The thermal dissociation of the excited state ($2p$) in the polarization
well adjusted to the ground state ($1s$) is
$(E_{\rm th})_{2p}^{1s}=\vert E_{2p,\rm min}^{1s}\vert$.

The elastic energy of the polarization field is
\begin{eqnarray}
(E_{\rm polariz})_{2p}^{1s}   =
{\lambda_{2p} \over 1+ \lambda_{2p}} + \sum_{\bf k} \vert V_{k}\vert ^2 \vert \rho_{\bf k}^{1s}\vert ^2,
\label{POLAR}\end{eqnarray}
which is similar to that for the $1s$ state. The first term in the right-hand
side of this equation is by two orders of
magnitude smaller  than the second one, therefore in both cases
\begin{eqnarray}
E_{\rm polariz}   \approx
\sum_{\bf k} \vert V_{k}\vert ^2 \vert \rho_{\bf k}^{1s}\vert ^2.
\label{POLAR1}\end{eqnarray}
The electron (hole) eigenenergy, corresponding to the non-relaxed excited
state $2p$ of a bound polaron, is
\begin{eqnarray}
(E_c)_{2p}^{1s}= E_{2p,\rm min}^{1s} - E_{\rm polariz},
\label{EE1}\end{eqnarray}
and the minimal energy, which would be absorbed during the photoionization
of a bound polaron from its non-relaxed excited state $2p$,
$(E_{\rm opt})_{2p}^{1s}=\vert (E_c)_{2p}^{1s} \vert$, takes the form
\begin{eqnarray}
(E_{\rm opt})_{2p}^{1s}=(E_{\rm th})_{2p}^{1s} + E_{\rm polariz}.
\label{PPP}\end{eqnarray}

Calculations for large polarons were performed with the following
parameters. (i) For MgO, $\hbar\omega_{\rm{\tiny LO}}=0.09$eV,
$\varepsilon_{\infty}$=2.96 and $\varepsilon_0$=9.86 \cite{S81},
$m_b=2.77m_0$ \cite{XC91}, where $m_0$ is the
bare electron mass; according to the definition (\ref{2}) $\alpha = 4.83$,
whereas the calculation by Eq.\,({\ref1}) gives $\beta = 4.14$.
(ii) For $\alpha$-Al$_2$O$_3$,
$\hbar\omega_{\rm{\tiny LO}}=0.07$eV\footnote{This value
is suggested in Ref. \cite{TF70} as an average of the phonon energies
corresponding to three modes which provide the strongest contributions to the
dielectric dispersion. This average value is in good agreement with the
more recent data \cite{BS75}, where the measured optical phonon energy
range from 0.052eV to 0.108eV.},
$\varepsilon_{\infty}$=3.1 and $\varepsilon_0$=9.0 \cite{TF70}.
Estimate of the hole effective mass in corundum from the experimental value
$E_{\rm opt}$ = 2.58 eV gives $m_h=4.0 m_0$, and consequently,
$\alpha = 5.41; \beta = 5.66$.

Atomistic calculations for bound hole polarons were performed using the
quantum chemical method of the  Intermediate Neglect of the Differential
Overlap (INDO) \cite{7} modified for ionic solids \cite{8}. This is based
on the Hartee-Fock formalism and allows self-consistent calculations of
the atomic and electronic structure of pure and defective crystals, that
is we are able to simulate polarons making no {\it a priori} assumptions about
their  geometrical structure. The INDO method was successfully applied
for the study of many defects in oxide crystals (see a review article
\cite{9} and references
therein). Quantum cluster is embedded into the electrostatic field of the
infinite non-point crystalline lattice. To model a hole polaron, we remove
one electron from the cluster and allow all atoms to relax to the minimum
of the total energy which gives us the atomic structure of a polaron. The
analysis of a relevant final charge density distribution characterizes the
polaron's electronic structure. Calculations of the difference of total
energies for the ground and excited states with a fixed lattice geometry
permit, in principle, to find the activation energy $E_{opt }$
of a charge transfer process, when a hole after
optical stimulation hops between O$^{2-}$ ions. In its turn, the thermal
dissociation energy $E_{\rm th}$ is calculated as the energy gain due to atomic
displacements induced by a hole. Results of the INDO method are
illustrated in Fig. 2.

In MgO, calculations were done for 125-atom clusters, modelling 9 spheres
of atoms surrounding a cation vacancy in the coordinate origin. The
basis set is the same as in our previous MgO studies \cite{10}. In corundum,
65-atom stoichiometric clusters containing 13 formula units were used.
The INDO parameters and basis set remain also the same as in previous
$\alpha$-Al$_2$O$_3$ calculations \cite{11}.
Both our microscopic calculations and ESR data \cite{2} demonstrate that a hole is
well localized by a single oxygen atom, forming O$^-$ ion considerably
displaced from its regular lattice site.
Practically, it is difficult to achieve a
convergence in the self-consistent-field calculations for the
delocalized (band) states necessary for calculations of the
photoionization energy. This is also the case in our calculations
for MgO and corundum. Neither we are aware of the relevant experimental data
on optical ionization energies of these polarons.

Results of calculations for the V$^-$ centers in MgO by theories CTAC,
CTSC and INDO and the available experimental data are collected in Table I.

The thermal dissociation energy $E_{\rm th}$ obtained in the continuum
theory is close to the experimental value; the difference of about 10\%
between them seems to be within the limits of the error of measurement. The INDO method gives the value of
$E_{\rm th}$ which is also close to the experimental data.
The energy $E_{\rm opt}$ calculated in the continuum theory
and within the INDO approach are both in a good agreement with experiment.
It is worth noting that CTAC better compares with experiment than CTST.

A detailed discussion of the optical properties of the V$^-$ centers in MgO
has been presented in terms of small-radius polarons by Schirmer \cite{12}.
Theory predicts the ratio of the optical to thermal
{\it reorientation} energy to be four, and the ratio to polarization energy to be two.
The latter relation is also supported by our CTSC calculations (Table I).
A very qualitative discussion of the
large-radius vs small-radius hole polarons in corundum has been done
in Ref. \cite{17}.

As distinct from the case of MgO, calculation of the polaron energy spectra
in corundum lacks for some material parameters.  The band mass $m_b$ has not been measured;
the value $\hbar\omega_{\rm{\tiny LO}}=0.07$eV \cite{TF70} is not fully reliable;
the thermal dissociation energy for the ground state of the
$V_{Mg}$ center is not known. The value of 0.7 eV from Table II is, in fact, the
activation energy of a free small polaron \cite{16}, which can be only conventionally
considered as $E_{\rm th}$. It is worth noting, that the shell-model calculation underestimates
the energy $E_{\rm th}$, resulting with 0.56 eV \cite{15}.
The energy  $E_{\rm th}$ calculated by using the continuum theory is close to
the respective result of the INDO approach; the both of them
differ from the experimental value.

In conclusion, we have shown that the continuum arbitrary-coupling theory (CTAC)
of large bound polarons qualitatively and quantitatively describes
the energy spectrum and transition energies for the impurity centers in
oxide compounds MgO and $\alpha$-Al$_2$O$_3$.
The thermal dissociation
energies obtained in the framework of CTAC are rather close to the results of
INDO. The CTAC has the advantage that it allows to describe the impurity
energy spectrum in detail. The surprising fact, that two essentially different
approaches have led to close results, seems to be indicative of
a dualistic nature of the bound polarons, which combine a {\it macroscopic}
scale of polarization changes with {\it microscopic} (atom-like) charge carrier states.
The efficiency of the proposed in our letter
macroscopic approach to the bound polarons in the aforementioned oxides is
probably due to
the fact, that the specific features of the small-radius (atom-like) states
of charge carriers do not appreciably influence the distribution of the
polarization field that is induced by themselves.

\acknowledgements

This work has been supported by the GOA BOF UA 2000, IUAP,
F.W.O.-V. projects No. G.0287.95, 9.0193.97 and WO.0025.99N (Belgium);
through DAAD
via a fellowship to E.A.K. at the Osnabr\"uck  University.
E.A.K. thanks the Scientific Research Community ``Low Dimensional Systems''
W.O.G. 0073.94N of the F.W.O.-Vlaanderen (Belgium) and E.P.P. acknowledges
the PHANTOMS Research Network and IUAP for the financial support of his visits to
the U.I.A.

\begin{figure}
{\large\bf
\centerline{
\setlength{\unitlength}{0.1bp}
\special{!
/gnudict 40 dict def
gnudict begin
/Color false def
/gnulinewidth 5.000 def
/vshift -33 def
/dl {10 mul} def
/hpt 31.5 def
/vpt 31.5 def
/M {moveto} bind def
/L {lineto} bind def
/R {rmoveto} bind def
/V {rlineto} bind def
/vpt2 vpt 2 mul def
/hpt2 hpt 2 mul def
/Lshow { currentpoint stroke M
  0 vshift R show } def
/Rshow { currentpoint stroke M
  dup stringwidth pop neg vshift R show } def
/Cshow { currentpoint stroke M
  dup stringwidth pop -2 div vshift R show } def
/DL { Color {setrgbcolor [] 0 setdash pop}
 {pop pop pop 0 setdash} ifelse } def
/BL { stroke gnulinewidth 2 mul setlinewidth } def
/AL { stroke gnulinewidth 2 div setlinewidth } def
/PL { stroke gnulinewidth setlinewidth } def
/LTb { BL [] 0 0 0 DL } def
/LTa { AL [1 dl 2 dl] 0 setdash 0 0 0 setrgbcolor } def
/LT0 { PL [] 0 1 0 DL } def
/LT1 { PL [4 dl 2 dl] 0 0 1 DL } def
/LT2 { PL [2 dl 3 dl] 1 0 0 DL } def
/LT3 { PL [1 dl 1.5 dl] 1 0 1 DL } def
/LT4 { PL [5 dl 2 dl 1 dl 2 dl] 0 1 1 DL } def
/LT5 { PL [4 dl 3 dl 1 dl 3 dl] 1 1 0 DL } def
/LT6 { PL [2 dl 2 dl 2 dl 4 dl] 0 0 0 DL } def
/LT7 { PL [2 dl 2 dl 2 dl 2 dl 2 dl 4 dl] 1 0.3 0 DL } def
/LT8 { PL [2 dl 2 dl 2 dl 2 dl 2 dl 2 dl 2 dl 4 dl] 0.5 0.5 0.5 DL } def
/P { stroke [] 0 setdash
  currentlinewidth 2 div sub M
  0 currentlinewidth V stroke } def
/D { stroke [] 0 setdash 2 copy vpt add M
  hpt neg vpt neg V hpt vpt neg V
  hpt vpt V hpt neg vpt V closepath stroke
  P } def
/A { stroke [] 0 setdash vpt sub M 0 vpt2 V
  currentpoint stroke M
  hpt neg vpt neg R hpt2 0 V stroke
  } def
/B { stroke [] 0 setdash 2 copy exch hpt sub exch vpt add M
  0 vpt2 neg V hpt2 0 V 0 vpt2 V
  hpt2 neg 0 V closepath stroke
  P } def
/C { stroke [] 0 setdash exch hpt sub exch vpt add M
  hpt2 vpt2 neg V currentpoint stroke M
  hpt2 neg 0 R hpt2 vpt2 V stroke } def
/T { stroke [] 0 setdash 2 copy vpt 1.12 mul add M
  hpt neg vpt -1.62 mul V
  hpt 2 mul 0 V
  hpt neg vpt 1.62 mul V closepath stroke
  P  } def
/S { 2 copy A C} def
end
}
\begin{picture}(2160,1943)(0,0)
\special{"
gnudict begin
gsave
50 50 translate
0.100 0.100 scale
0 setgray
/Helvetica findfont 100 scalefont setfont
newpath
-500.000000 -500.000000 translate
LTa
600 251 M
0 1641 V
LTb
600 251 M
63 0 V
1314 0 R
-63 0 V
600 661 M
63 0 V
1314 0 R
-63 0 V
600 1072 M
63 0 V
1314 0 R
-63 0 V
600 1482 M
63 0 V
1314 0 R
-63 0 V
600 1892 M
63 0 V
1314 0 R
-63 0 V
600 251 M
0 63 V
0 1578 R
0 -63 V
875 251 M
0 63 V
0 1578 R
0 -63 V
1151 251 M
0 63 V
0 1578 R
0 -63 V
1426 251 M
0 63 V
0 1578 R
0 -63 V
1702 251 M
0 63 V
0 1578 R
0 -63 V
1977 251 M
0 63 V
0 1578 R
0 -63 V
600 251 M
1377 0 V
0 1641 V
-1377 0 V
600 251 L
LT0
1674 1729 M
180 0 V
600 1535 M
28 -14 V
27 -17 V
28 -15 V
27 -18 V
28 -19 V
27 -19 V
28 -18 V
27 -18 V
28 -18 V
27 -20 V
138 -87 V
138 -79 V
137 -71 V
138 -63 V
138 -56 V
138 -49 V
137 -31 V
138 -25 V
LT1
1674 1629 M
180 0 V
600 1401 M
28 -6 V
27 -9 V
28 -12 V
27 -13 V
28 -14 V
27 -16 V
28 -16 V
27 -17 V
28 -16 V
27 -17 V
138 -86 V
138 -79 V
137 -70 V
138 -62 V
138 -54 V
138 -47 V
137 -34 V
138 -49 V
LT2
1674 1529 M
180 0 V
600 1199 M
28 9 V
27 3 V
28 -1 V
27 -2 V
28 -8 V
27 -9 V
28 -12 V
27 -13 V
28 -15 V
27 -15 V
138 -81 V
138 -73 V
137 -68 V
138 -58 V
138 -50 V
138 -44 V
137 -37 V
138 -33 V
LT3
1674 1429 M
180 0 V
600 923 M
28 27 V
27 22 V
28 13 V
27 10 V
28 2 V
27 0 V
28 -5 V
27 -7 V
28 -9 V
27 -12 V
138 -67 V
138 -70 V
137 -61 V
138 -53 V
138 -43 V
138 -37 V
137 -31 V
138 -27 V
LT4
1674 1329 M
180 0 V
600 621 M
28 40 V
27 30 V
28 26 V
27 15 V
28 12 V
27 4 V
28 -1 V
27 -4 V
28 -7 V
27 -9 V
138 -58 V
138 -58 V
137 -48 V
138 -38 V
138 -31 V
138 -25 V
137 -20 V
138 -17 V
LT6
1674 1229 M
180 0 V
600 522 M
28 36 V
27 25 V
28 16 V
27 8 V
28 0 V
27 -4 V
28 -7 V
27 -10 V
28 -10 V
27 -12 V
138 -58 V
138 -45 V
137 -34 V
138 -24 V
138 -19 V
138 -15 V
137 -12 V
138 -10 V
stroke
grestore
end
showpage
}
\put(1614,1229){\makebox(0,0)[r]{{\normalsize 1}}}
\put(1614,1329){\makebox(0,0)[r]{{\normalsize 2}}}
\put(1614,1429){\makebox(0,0)[r]{{\normalsize 4}}}
\put(1614,1529){\makebox(0,0)[r]{{\normalsize 6}}}
\put(1614,1629){\makebox(0,0)[r]{{\normalsize 8}}}
\put(1614,1729){\makebox(0,0)[r]{{\normalsize $\alpha$=10}}}
\put(1288,51){\makebox(0,0){$\beta$}}
\put(100,1071){%
\special{ps: gsave currentpoint currentpoint translate
270 rotate neg exch neg exch translate}%
\makebox(0,0)[b]{\shortstack{$\kappa=E_{{\rm opt},1s}/E_{{\rm th},1S}$}}%
\special{ps: currentpoint grestore moveto}%
}
\put(1977,151){\makebox(0,0){10}}
\put(1702,151){\makebox(0,0){8}}
\put(1426,151){\makebox(0,0){6}}
\put(1151,151){\makebox(0,0){4}}
\put(875,151){\makebox(0,0){2}}
\put(600,151){\makebox(0,0){0}}
\put(540,1892){\makebox(0,0)[r]{3.0}}
\put(540,1482){\makebox(0,0)[r]{2.5}}
\put(540,1072){\makebox(0,0)[r]{2.0}}
\put(540,661){\makebox(0,0)[r]{1.5}}
\put(540,251){\makebox(0,0)[r]{1.0}}
\end{picture}}
}
\vskip 1cm
\caption{Ratio $\kappa (\alpha,\beta)$ of the photoionization and thermal
dissociation
energies as a function of the strength of the Coulomb interaction $\beta$
for a large polaron bound to an impurity
with different values of the Fr\"ohlich coupling
constant $\alpha$. Note that for large constants $\alpha$, the calculated
values $\kappa (\alpha,0)$ tend to 3, in conformity with the limiting
result of the adiabatic theory \protect\cite{5}.}
\end{figure}
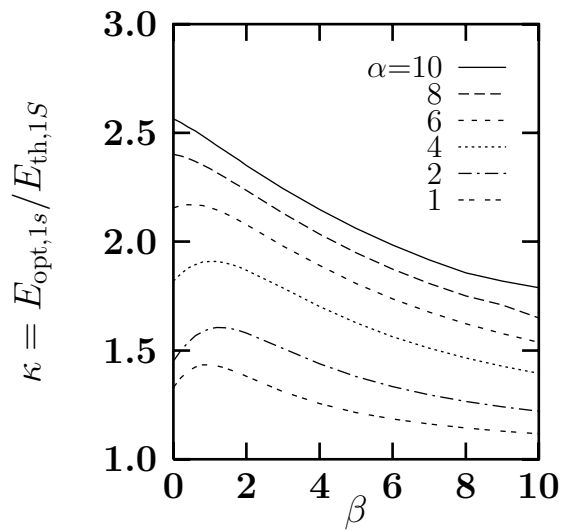

\begin{figure}
\vbox to 0.5cm{}
\caption{Visualisation of the atomic structure of the V$^-$ center in MgO
(a) and
V$_{Mg}$ center in corundum (b) as calculated by means of the INDO method.
In the V$^-$ center five nearest neighbour O atoms move outwards from the
Mg vacancy by about 1\% $a_0$ (the lattice parameter), whereas the O atom
with a hole
(shaded) is attracted to the Mg vacancy by 3\% $a_0$ which makes the
dominant
contribution to the energy gain due to the lattice relaxation.}
\end{figure}

\begin{table}\caption{Theoretical and experimental values of the
$ E_{\rm th}$  and  $E_{\rm opt}$  for a polaron bound to the Mg
vacancy (the $V^{-}$-center) in MgO. CTAC, CTSC, INDO stand for
continuum arbitrary-coupling
theory, continuum strong-coupling theory, and microscopic quantum-chemical approach,
correspondingly; experimental values are given in the last column.
$\hbar\omega_{ex}$ is a transition energy between the ground and excited states.
Sign ``$-$'' indicates that the respective quantities were not calculated;
indexes $g$ and $e$ label, correspondingly, the ground and excited states
(those states are $1s$ and $2p$ in CTAC).}

\bigskip
\begin{center}
\begin{tabular}{|l|l|l|l|l|}\hline
 Energies (eV) &  CTAC          & INDO       & CTSC & experiment \\ \hline
$E_{\rm th}$ &     1.24 & 1.6 & 1.15     &    1.4 \cite{4}         \\  \hline
$E_{\rm opt}$          &   2.2     & 2.2 & 2.13       & 2.3 \cite{12}\\  \hline
$(E_{\rm opt})^g_e$     &  1.13  &  $-$        &  0.96   &  \\ \hline
$\hbar\omega_{ex} = E_{\rm opt} - (E_{\rm opt})^g_e$     &1.07&  $-$     &1.16 & \\ \hline
$E_{\rm polariz} $& 0.9 &$-$& 0.97  &           \\ \hline
$\kappa $& 1.77 & 1.38 & 1.85 &    1.64   \\ \hline
\end{tabular}
\smallskip
\end{center}
\end{table}

\begin{table}\caption{ Same as Table 1
for the $V_{Mg}$-center in $\alpha$-Al$_2$O$_3$}

\bigskip
\begin{center}
\begin{tabular}{|l|l|l|l|}\hline
 Energies (eV) &  CTAC          & INDO       & experiment \\ \hline
$E_{\rm th}$ &     1.22 &  1.26     &  0.68  \cite{13}; 0.7\cite{14}         \\  \hline
$E_{\rm opt}$          &   (2.58)$^*$ & $-$       & 2.58 \cite{13}\\  \hline
$(E_{\rm opt})^g_e$          &  1.15  &  $-$      &  \\ \hline
$\hbar\omega_{ex} = E_{\rm opt} - (E_{\rm opt})^g_e$     &1.43&   $-$   & \\ \hline
$E_{\rm polariz} $& 1.36 & $-$ &           \\  \hline
$\kappa $& 2.11 & $-$ &  3.68   \\ \hline
\end{tabular}
\smallskip
\end{center}
{\small
\noindent  $^{*}$
The value of the energy $E_{\rm opt} = 2.58 eV$ was used
to estimate the effective mass of a hole: $m_b=4m_0$.
}
\end{table}


\begin{references}
\bibitem[*]{A1} Also at: Universiteit Antwerpen (RUCA), Groenenborgerlaan 171,
B-2020 Antwerpen, Belgium and
Technische Universiteit Eindhoven, P. O. Box 513,
5600 MB Eindhoven, The Netherlands.
\bibitem[\sharp]{A2} Permanent address: Laboratory of Multilayer Structure Physics,
Department of Theoretical Physics,
State University of Moldova,
Strada A. Mateevici, 60,
MD-2009 Kishinev, Moldova.
\bibitem[\dag]{A3}
Also at: Institute of Solid State Physics, University of Latvia,
Kengaraga 8, LV-1063 Riga, Latvia.

\bibitem{1} {A. S. Alexandrov and N. Mott},
{\it Polarons and Bipolarons}, World Scientific, Singapore,
{1996}.
\bibitem{2} {J. T. Devreese}, Polarons, in:
{\it Encyclopedia of Applied Physics} {\bf 14 }, 383 (1996).
\bibitem{3} {A. L. Shluger and A. M. Stoneham}, {J. Phys.: Cond.
Mat.} {\bf 5}, 3049 (1993).
\bibitem{4} {Y. Chen and M. M. Abraham},
 {J. Phys. Chem. Sol.} {\bf 51}, 747 (1990).
\bibitem{5} {S. I. Pekar},
{\it Untersuchungen \"uber die Elektronentheorie der Kristalle},
 Akademie-Verlag, Berlin, 1954.
\bibitem{L73} {L. F. Lemmens, J. De Sitter, and J. T. Devreese},
 { Phys. Rev. B} {\bf  73}, 2717 (1973).
\bibitem{E65}
{ J. Devreese and R. Evrard},
 { Phys. Lett.} {\bf  11}, 278 (1964);
{ R. Evrard},
 { Phys. Lett.} {\bf  14}, 295 (1965).
\bibitem{A99} { A. S. Alexandrov and P. E. Kornilovitch},
 {Phys. Rev. Lett.} {\bf 82}, 807 (1999).
\bibitem{6} { J. Devreese, R. Evrard, E. Kartheuser, and F. Brosens},
 { Solid State Commun.} {\bf  44}, 1435 (1982).
\bibitem{S81} { M. J. L. Sangster},  {Phil. Mag. B}
{\bf 43}, 597 (1981).
\bibitem{XC91} {Yong-Nian Xu  and W. Y. Ching},  {Phys. Rev. B}
{\bf 43}, 4461 (1991).
\bibitem{TF70} {K. K. Thornber and R. P. Feynman},  {Phys. Rev. B}
{\bf 1}, 4099 (1970).
\bibitem{BS75} {H. Bialas and H. J. Stolz},  {Z. Physik B}
{\bf 21}, 319 (1975).
\bibitem{7} { J. A. Pople and D. L. Beverage},
{\it Approximate Molecular Orbital Theory},
McGraw Hill, New York, 1970.
\bibitem{8} { E. V. Stefanovich, E. Shidlovskaya, and A. L. Shluger},
 { Phys. Stat. Sol. B} {\bf  160}, 529 (1990);
{ A. L. Shluger and E. V. Stefanovich},  {Phys. Rev. B} {\bf 42}, 9664
(1990).
\bibitem{9} { E. A. Kotomin and A. I. Popov},
 {Nucl. Inst. Meth. B} {\bf 141}, 1 (1998).
\bibitem{10} { E. A. Kotomin, M. M. Kuklja, R. Eglitis, and A. I. Popov},
 {Mater. Sci. Eng. B} {\bf 37}, 212 (1996);
 {Comput. Mater. Sci.} {\bf  5}, 298 (1996).
\bibitem{11} { P. W. M. Jacobs, E. A. Kotomin, A. Stashans, E. V.
Stefanovich, and I. A. Tale},
 {J Phys: Cond. Matt.} {\bf  4}, 7531 (1992);
{ P. W. M. Jacobs and E. A. Kotomin},
 {Phys. Rev. Lett.} {\bf  69}, 1411 (1992).
\bibitem{12}  { O. F. Schirmer},
 { Z. Physik B} {\bf 24}, 235 (1976).
\bibitem{13} { H. A. Wang, C. H. Lee, F. A. Kroeger, and R. T. Cox},
 {Phys. Rev. B} {\bf 27 }, 3821 (1983).
\bibitem{14} { R. T. Cox}, in: {\it Proc. 1980 Annual Conf. of
European Phys. Soc. (Cond. Matt. Div.)}, Antwerpen, 1980, p. {318}.
\bibitem{15} { C. R. A. Catlow, R. James, W. C. Mackrodt, and R. F. Stewart},
 { Phys. Rev. B} {\bf 25}, 1006 (1982).
\bibitem{16} { E. A. Kotomin, I. A. Tale, V. G. Tale, P. Butlers, and P.
Kulis}, {J Phys: Cond. Matt.} {\bf 1}, 6777 (1989).
\bibitem{17} { E. A. Colbourn and W. C. Mackrodt},
 { Solid State Commun.} {\bf  40}, 265 (1981).
\end{references}
\end{document}